\documentclass[aip,jcp,reprint,floatfix,onecolumn]{revtex4-1}

\usepackage{graphicx}              
\usepackage{bm}                    
\usepackage{amsmath}

\def\beq{\begin{equation}}
\def\eeq{\end{equation}}

\def\det{\text{det}}
\def\tr{\text{tr}}
\def\A{A}

\def\P{P}

\def\Al{B}

\def\aa{A^{-1}\tilde{A}}
\def\wdet{D}

\begin{document}

\title{Simple formalism for efficient derivatives and multi-determinant expansions in quantum Monte Carlo}
\author{Claudia Filippi}
\email{c.filippi@utwente.nl}
\affiliation{MESA+ Institute for Nanotechnology, University of Twente, P.O. Box 217, 7500 AE Enschede, The Netherlands}
\author{Roland Assaraf}
\email{assaraf@lct.jussieu.fr}
\affiliation{Laboratoire de Chimie Th\'eorique, CNRS, UMR 7616, Universit\'e Pierre et Marie Curie Paris VI, Case 137, 4 Place Jussieu, 75252 Paris Cedex 05, France}
\author{S. Moroni}
\email{moroni@democritos.it}
\affiliation{CNR-IOM DEMOCRITOS, Istituto Officina dei Materiali, and SISSA Scuola Internazionale Superiore di Studi Avanzati, Via Bonomea 265, I-34136 Trieste, Italy} 

\begin{abstract}
We present a simple and general formalism to compute efficiently the derivatives of a multi-determinant Jastrow-Slater wave function, the local energy, the interatomic forces, and similar  quantities needed in quantum Monte Carlo. 
Through a straightforward manipulation of matrices evaluated on the occupied and virtual orbitals, we obtain an 
efficiency equivalent to algorithmic differentiation in the computation of the interatomic forces and the optimization of the orbital 
paramaters. Furthermore, for a large multi-determinant expansion, the significant computational gain recently reported for the calculation 
of the wave function is  here improved and  extended to all local properties in both all-electron and pseudopotential calculations.

\end{abstract}

\maketitle

\section{Introduction}

In the application of quantum Monte Carlo (QMC) methods to electronic systems in real space~\cite{Foulkes01,Kolorenc11}, one 
computes expectation values of random 
variables depending on $\psi ({\bf R})$ where $\psi$ is a variational ansatz of the exact wave function, and ${\bf R}=({\bf r}_1 \dots {\bf r}_N)$ 
are the coordinates of the $N$ electrons.  The total energy is for instance estimated as the expectation value of the local energy 
${\rm E}_{\rm L}=\hat{H}\psi/\psi$, where $\hat{H}$ is the Hamiltonian of the system. Other examples are the derivatives of $\psi$ and ${\rm E}_{\rm L}$  
with respect to the atomic coordinates or the variational parameters, which are needed to evaluate the interatomic forces or to optimize
the wave function $\psi$, respectively.
It is very important to compute these quantities efficiently because of their large number, typically $O(N)$ or $O(N^2)$, and also because 
they must be calculated for the many steps of the sampling process needed to collect significant statistics on the quantity of interest.

Here, we propose  a general, simple and efficient method to compute these properties for the most common ansatz of $\psi$ found in the literature, 
namely, a sum of $N_e+1$ Slater determinants times a Jastrow correlation factor $J ({\bf R})$,
\begin{equation}
\psi ({\bf R}) = J ({\bf R}) (D_0 + \sum_{i=1}^{N_e} c_i D_i) \,,
\label{cijas}
\end{equation}
where $D_0$ is a reference determinant (the Hartree-Fock solution, for example) of spin-orbitals (one-body functions depending on the position and 
the spin) and $D_i$ are excited determinants.  The formulation we propose here relies on the fact that one-body operators and derivatives can be written 
using the same compact expression, namely, the trace of the product of two matrices, when acting on a one-determinant Jastrow-Slater wave function.  
Consequently, derivatives, local quantities, and derivatives of local quantities are easy to obtain for the reference $JD_0$ and can be very 
simply and efficiently updated when $D_0$ is replaced by an excited determinant $D_i$. 
In practice, the method requires only the calculation of molecular orbitals and their derivatives with respect to some parameter related to 
the quantity being computed (e.g.\ the local energy or the derivative of the local energy) for all positions ${\bf r}_i$ of the electrons.  This 
information is stored in rectangular matrices of size $N \times N_{\rm orb}$ where $N_{\rm orb}$ is the total number of orbitals (occupied in the
reference plus virtual in case of a multi-determinant expansion). 
Such one-body quantities are very simple to code and already available in many QMC codes.  One has then to apply the few formulas we develop 
that involve  inverses and products of selected square submatrices.  Because these formulas are simple and common to all the properties introduced above, this 
method requires a minimal programming effort.

Our theoretical framework is very efficient in the regime of small and large $N_e$. In case of a single-determinant wave function so often used 
in QMC calculations and for small $N_e$, our formulas can for instance be used to achieve great computational savings in the evaluation of 
first-order derivatives such as the $3N_{\rm atoms}$ internuclear forces using zero-variance estimators and the space-warp 
transformation~\cite{Umrigar89,Filippi00}.
We recover in fact the same scaling of $O(N^3)$ which was obtained in Ref.~\onlinecite{Sorella10} with the use of algorithmic 
differentiation (AD). Here, however, we do not need to employ AD since we have at our disposal a very simple and transparent formula for the 
derivative of the energy.  A similar favorable scaling is also obtained in the computation of the derivatives needed to optimize all orbital 
parameters in the determinantal component of the wave function.

In the large $N_e$ regime, we provide a very compact formula for the computation of any local quantity and its derivatives evaluated for an excited 
determinant, $D_i$, by exploiting that $D_i$ differs from $D_0$ by a few orbital excitations.
In the calculation of $\psi$, we know from the work by Clark {\sl et al.}~\cite{Clark11} that, once $D_0$ has been computed, $D_i$ can be updated using 
matrices of order $k$ where $k$ is the order of the excitation.
 Here, because in our formulation all the properties introduced above are treated on an equal footing,  the favorable asymptotic  scaling of $O(k^3 N_e)$, 
instead of $O(N^3 N_e)$ or $O(N^2 N_e)$, obtained~\cite{Clark11} in the calculation of $\psi$, $\nabla_i \psi$ and $\Delta_i \psi$ 
applies to all properties including ${\rm E}_{\rm L}$ and its derivatives.  
For  $\nabla_i \psi$ and $\Delta_i \psi$, we also avoid  an additional $O(N N_s)$ term, where $N_s$ is the number of active single excitations.
Importantly, it holds in all-electron and  pseudopotentials calculations alike.
We stress that the formulas we propose here are general and that we do not employ strategies which exploit
further assumptions on the wave function $\psi$ (e.g.\ the possible equality of some determinants of a given spin in the expansion~\cite{Scemama15}), 
which could of course be introduced to further improve the scaling. 
 
The remainder of this paper is organized as follows: In Section~\ref{O_local}, we introduce the formulas for a local operators acting on a single
determinant and its derivatives, which are extended to the multi-determinantal case in Section~\ref{mexc}. 
In Section~\ref{jaspseB}, we present further details on how the expressions are modified in the presence of the Jastrow factor 
or non-local pseudopotentials and give numerical demonstration of our formulation in Section~\ref{numexe}. The formulas for the second
derivative of an excited determinant are given in Section.~\ref{Sec_2nd_mult}.

\section{Derivatives and one-body operators}
\label{O_local}

We begin with a single (Hartree-Fock) Slater determinant with occupied orbitals, $\phi_1 \dots \phi_N$, and denote it as
\begin{equation}
\wdet({\bf R}) =  |\phi_1   \phi_2  \dots  \phi_N| = \sum_P  (-1)^P \phi_{1}({\bf r}_{P(1)})  \phi_{2}({\bf r}_{P(2)}) \dots \phi_N ({\bf r}_{P(N)}) =  
\det (A)\,,
\end{equation}
with the Slater matrix $A$,
\begin{equation}
A_{ij} = \phi_j ({\bf r}_i)\,.
\end{equation}
The orbitals and the electrons correspond respectively to the columns and the lines.

Many important quantities like the local energy, the drift or the internuclei forces involve the derivative  of a Slater determinant with respect to some parameter $\lambda$.
The following identity will be the basis of all subsequent developments 
\begin{equation}
\frac{ d\ln D}{d\lambda} = \tr( A^{-1} B) \text{ \ \ where \ } B = \frac{dA}{d\lambda}\,,
\label{derivtrace}
\end{equation}
where the dependence of $A$ with respect to the parameter $\lambda$ is implicit.
For a proof, one can for example resort to simple chain rule and differentiate  with respect to the elements of $A$
\begin{equation}
d\ln D=\sum_{ij} \frac{d\ln D}{d A_{ij}} d A_{ij} = \tr(A^{-1}\,d A)\,.
\label{tr_deriv_det}
\end{equation}
The second equality comes from the expansion of the determinant in minors.
If $\lambda$ is the first coordinate of the first electron $x_1$, one obtains the corresponding  component of the drift 
\begin{equation}
\frac{1}{\wdet} \frac{\partial \wdet}{\partial x_1} =  \tr (A^{-1} B) \text{\  \ with \ } \Al = \frac{\partial  A}{\partial x_1}\,,
\label{drift11}
\end{equation}
where the matrix $B$ is zero with the exception of the first row.
If one is interested in computing the interatomic forces, one needs to evaluate the derivative of $\psi$ with respect to the
atomic coordinates and can employs the same formula with $\Al = {\partial  A}/{\partial R_a}$ where $R_a$ is an atomic coordinate.
Derivatives of a Slater determinant with respect to any variational parameter of the orbitals are also useful for optimization purposes.

Importantly, the application of a one-body operator to a Slater determinant can also be written as a first-order derivative for an appropriate choice of 
the matrix $B$. This observation will be at the core of the very efficient computation of local quantities and their gradients for 
single- and large multi-determinantal wave functions.
To show this, we consider the one-body operator
\begin{equation}
\hat{O} = O({\bf r}_1) + \dots O({\bf r}_N)\,,
\end{equation}
where $O({\bf r}_i)$ is an operator which acts only on a function of ${\bf r}_i$. 
Applying the operator to the determinant as
\begin{equation}
\hat{O} \wdet =  \sum_P  (-1)^P  (O({\bf r}_1) + \dots O({\bf r}_N))
\phi_{1}({\bf r}_{P(1)})  \phi_{2}({\bf r}_{P(2)}) \dots \phi_N ({\bf r}_{P(N)})  \,,
\end{equation}
and expanding the product inside the sum, we have  
\begin{equation}
\hat{O} \wdet = |(O \phi_1)  \phi_2 \dots \phi_N| + |\phi_1  (O\phi_2) \dots \phi_N| + \dots   |\phi_1  \phi_2 \dots (O \phi_N) |\,,
\label{omono}
\end{equation}
which is the sum of all mono-excitations obtained by replacing in turn each orbital $\phi_i$ with $O \phi_i$.
It is easy to check that 
\begin{equation}
\frac{\hat{O}\wdet}{\wdet} = \frac{d}{d\lambda} \ln  \det (\A + \lambda B) =  \tr (\A^{-1}  B) \,,
\label{dettrace0}
\end{equation}
where the derivative is taken at $\lambda=0$ and
\begin{equation}
B_{ij} = (O\phi_j)({\bf r}_i) \,.
\end{equation}
To prove the first identity in  Eq.~(\ref{dettrace0}), we just have to perform the derivative of 
\begin{equation}
\det(A+\lambda B) = |(\phi_1 + \lambda O \phi_1) (\phi_2 + \lambda O \phi_2) \dots | 
\end{equation}
with respect to $\lambda$ and use the multi-linearity of the determinant. The second identity follows from Eq.~(\ref{derivtrace}).
A very important example is the kinetic energy operator $\hat{O}=\hat{T}$ with
\begin{equation}
O({\bf r}_i)  = -\frac{1}{2}\Delta_i   \equiv -\frac{1}{2} \left(\frac{\partial^2}{\partial x_i^2}+  
\frac{\partial^2}{\partial y_i^2} + \frac{\partial^2}{\partial z_i^2}\right) \,.
\end{equation}
Eq. (\ref{dettrace0}) then holds with the definition 
\begin{equation}
B_{ij} = -\frac{1}{2}\Delta A_{ij} =- \frac{1}{2} \Delta \phi_j ({\bf r}_i)\,.
\label{B_kin_det}
\end{equation}
In other words, the Laplacian can be written as a first-order derivative when applied to a Slater determinant. 

Expression (\ref{derivtrace}) or (\ref{dettrace0}) can be generalized to  wave functions including a Jastrow factor and to other operators like a non-local pseudopotential.
The corresponding $B$ matrices are easy   to write and will be given later.

\subsection{Second-order derivatives}
\label{2deriv_1det}

The compact trace expressions of a local quantity (Eq.~\ref{dettrace0}) offers the advantage that its derivative with respect to a parameter $\mu$ can be straightforwardly written as
\begin{eqnarray}
\frac{\partial}{\partial \mu}\frac{\hat{O}\wdet}{\wdet}  
={\rm tr}(A^{-1}\partial_\mu B-X\partial_\mu A)\,,
\label{dm_dettrace0}
\end{eqnarray}
where $\partial_\mu A$ and $\partial_\mu B$ are the matrices of the derivatives of the elements of $A$ and $B$, respectively, and the matrix
$X$ is defined as
\begin{eqnarray}
X=A^{-1}{B}A^{-1}\,.
\label{X_mat}
\end{eqnarray}
This can easily be shown by using $d(A^{-1})=-A^{-1}dA\,A^{-1}$ and the cyclic property of the trace.

Therefore, to compute the derivative of a local quantity with respect to many parameters, one evaluates and stores the matrix $X$ at a cost 
proportional to $N^3$ and then computes its derivatives with the cost for each parameter being 
at most $N^2$. For instance, in the computation of the derivatives of $E_{\rm L}$ with respect to the 
nuclear coordinates, this procedure allows the 
efficient computation of the forces with a cost per Monte Carlo step proportional to the one of the energy:  Since the number of 
atoms $N_{\rm atoms}$ scales linearly with the number of electrons, the cost of computing the forces at each Monte Carlo step is 
$N^3+N_{\rm atoms}\times N^2\sim N^3$.  

The same expression (Eq.~\ref{dm_dettrace0}) can be used in an orbital optimization run to efficiently compute the derivatives of the local energy with 
respect to the orbital variations. If we consider $\phi_i\to\phi_i+\mu\,\phi_j$ where $\phi_i$ is occupied in the original determinant and $\phi_j$ unoccupied,  the derivative of a local quantity with respect to $\mu$ is 
\begin{equation}
\frac{d}{d\mu} \frac{{\hat{O}}\wdet}{\wdet} = \tilde{M}_{ij}\,,
\label{devmono}
\end{equation}
where the derivative is taken at $\mu = 0$ and the rectangular matrix
\begin{equation}
\tilde{M} = A^{-1} \tilde{B} -X \tilde{A} = A^{-1} \tilde{B} -A^{-1}BA^{-1} \tilde{A} \,,
\label{deftildeM}
\end{equation}
is computed from the $\tilde{A}$ and $\tilde{B}$ rectangular extensions of $A$ and $B$ to the $N_{\rm virt}$ virtual orbitals.
To show this, we simply note that all elements of the matrices $\partial_\mu A$ and $\partial_\mu B$ are zero with 
the exception of the $i^{\rm th}$ row which contains $\phi_j$ and $O\psi_j$, respectively. Then, 
if the number of orbital variations is $N_s$ (which equals the number of determinants $\bar{D}$ created via single excitations $D\to D+\mu\bar{D}$), the cost of evaluating
the matrix $\tilde{M}$ is only $2N_s\times N$. In a standard implementation, one would 
instead compute the full inverse matrix of the corresponding mono-excitation to obtain for example the derivatives with respect to the electron positions 
in the kinetic energy with a cost proportional to $N_s\times N^2$.
Therefore, also in the optimization of the determinantal component as in the case of the interatomic forces, our formulation leads to total cost of estimating the needed quantities 
proportional to the cost of computing the energy, $N^3$, since $N_s$  grows at most like $N\times N_{\rm virt} \sim N^2$.

Thanks to Eq.~(\ref{derivtrace}), the same formula~(\ref{dm_dettrace0}) also applies to the second derivative $\partial_\mu\partial_\lambda \ln D$, where the matrix $B$ is then equal to $\partial_\lambda A$. The expression can then be cast in a form where $\lambda$ and $\mu$ enter symmetrically,
\begin{eqnarray}
\frac{\partial^2  \ln D } {\partial \mu \partial \lambda} 
={\rm tr}(A^{-1}\partial_\mu\partial_\lambda A-(A^{-1}\partial_\lambda A)(A^{-1}\partial_\mu A))\,,
\label{2deriv_sym}
\end{eqnarray}
Also in this case, however, if $\lambda$ and $\mu$ denote two sets of variables and one set, for instance $\{\lambda\}$, is significantly smaller than the other, it is computationally convenient to group the matrices differently and precompute the matrices $X_\lambda =A^{-1}\partial_\lambda A A^{-1}$, followed by the evaluation of the trace with the more numerous $\partial_\mu A$.

\section{Multiple excitations}
\label{mexc}

We now consider multiple excitations of the original Slater determinant and deduce all subsequent formulas from Eq.~(\ref{dettrace0}) 
where the derivative is taken at $\lambda=0$.

\subsection{Determinant of a multiple excitation}

If $k$ columns of $\A$ are  modified (in an excitation of order $k$), the new Slater determinant is
\begin{equation}
\bar{\wdet} = \det (\bar{A})\,.
\end{equation}
To compute $\bar{\wdet}$, we introduce the  projector $P$ on the space of columns which are  modified, namely,
a diagonal $N\times N$ matrix such that $P_{ii}=1$ if the orbital $i$ has been modified and zero otherwise.
For example, if only the first and third column of $\bar{A}$ and $A$ are different,
\begin{equation}
P = \left( \begin{array}{cccccc}
1 & 0 & 0 & 0 & \cdots \\ 
0 & 0 & 0 & 0 & \cdots \\ 
0 & 0 & 1 & 0 & \cdots \\
0 & 0 & 0 & 0 &\dots \\
\vdots & \vdots & \vdots & \vdots &  &\\
\end{array} \right)
\end{equation}
Then, since $\bar{A}-A = (\bar{A}-A)P$, we have 
\begin{eqnarray}
\det (\bar{\A}) 
 &  =&  \det(A + (\bar{A}-A)P) \nonumber\\ 
& = &  \det (\A) \det (1 +  \A^{-1} (\bar{\A}-\A) \P) \nonumber\\
&  = &  \det (\A) \det (1-\P +  \A^{-1} \bar{\A} \P)  \,.
\label{nop} 
  \end{eqnarray}
The matrix $1-\P +  \A^{-1} \bar{\A} \P$ is the identity in the $(N-k)\times(N-k)$ block projected by $1-P$ because all columns of $\A^{-1} \bar{\A} \P$ in the same block are zero. The determinant of the total matrix $1-\P +  \A^{-1} \bar{\A} \P$ is therefore the determinant of the remaining $k\times k$ block which is $\P \A^{-1} \bar{\A} \P$:
\begin{eqnarray}
\det (\bar{\A}) & = & \det (\A)  \det_P (\P \A^{-1} \bar{\A} \P) \,,
\label{detk}
 \end{eqnarray}
where the subscript $P$ is introduced to specify that the determinant is computed for the block identified by the projector $P$.
We note that this formula can also be proved using the determinant lemma \cite{Clark11}.

In practice,  once we have  computed $\A^{-1}$ and $\det(A)$, we can evaluate and store the matrix   $\A^{-1} \tilde{A}$ where $\tilde{A}$ is the rectangular extension of $A$ to the unoccupied orbitals.
For a $k^{th}$-order excitation, we only have to compute the determinant of 
 $\P \A^{-1}  \bar{\A}\P$ which is a simple submatrix of $A^{-1}\tilde{A}$ with  dimension $k\times k$. This submatrix is built 
by selecting the coefficients $(\aa)_{ij}$ such that   $i$ is the index of  a substituted orbital and $j$ is the index of an excited orbital.
For example, if $1 \to 11$ and $3 \to 15$ are the list of excitations,
the matrix is
\begin{equation}
PA^{-1}\bar{A}P = \left[\begin{matrix} (\aa)_{1,11} & (\aa)_{1,15}   \\ 
                      (\aa)_{3,11} & (\aa)_{3,15}                         
\end{matrix} \right]\,.
\end{equation}
Note that the first block composed of the $N$ first columns of $\A^{-1} \tilde{A}$ corresponds to occupied orbitals and is the identity matrix. It is never used and does not have to be stored. In practice, one needs to compute only the $N\times N_{\rm virt}$ submatrix where $N_{\rm virt}$ is the number of virtual orbitals present in the multi-determinant expansion.

\subsection{One-body operator applied to a $k^{\rm th}$-excited  determinant }
\label{O_multi-det}

Applying formula (\ref{dettrace0}) to a new determinant $\bar{\wdet}$, we have
\begin{eqnarray}
\frac{\hat{O} \bar{\wdet}} {\bar{\wdet}} = 
\tr (\bar{A}^{-1} \bar{B}) = 
\frac{d}{d\lambda}\ln \det (\bar{\A} + \lambda {\bar B})\,,
\label{onewdet}
\end{eqnarray}
where $\bar{A}$ is built with different orbitals and $\bar{B}$ is the corresponding new matrix $B$.
Note that, similarly to $\bar{A}-A$, the columns of $\bar{B}-B$ corresponding to non-excited orbitals are zero.
Using equation (\ref{detk}),
\begin{eqnarray}
 \tr (\bar{A}^{-1} \bar{B}) &= &
\frac{d}{d\lambda}\ln  [  \det (\A+\lambda B) \det_\P ( \P (\A+\lambda B)
^{-1} (\bar{\A}+\lambda \bar{B}) \P)]\,,
\nonumber
\end{eqnarray}
and computing the derivative at $\lambda=0$, we have
\begin{equation}
\tr (\bar{A}^{-1} \bar{B}) 
= \tr (\A^{-1} \Al) + \tr ( (\P \A^{-1} \bar{\A}\P)^{-1} P \bar{M}P )\,,
\label{multiplex}
\end{equation}
where we used again Eq.~(\ref{tr_deriv_det}) and defined 
\begin{eqnarray}
\bar{M} & \equiv & \A^{-1} \bar{\Al}-X  \bar{\A} 
\end{eqnarray}
The matrix $X$ is given in Eq.~(\ref{X_mat}) and  
$(\P \A^{-1}\bar{A} \P)^{-1}$ is such that $(\P  \A^{-1}\bar{A} \P)^{-1} (\P \A^{-1}\bar{A} \P) = \P$. We have omitted the index $P$ in the trace above since $\tr_P$ yields the same result as the complete trace, which runs over additional zero matrix elements.
We recall that the same expression~(\ref{multiplex}) also applies to the logarithmic derivative $\partial_\lambda \bar{D}$ where 
$B=\partial_\lambda A$ and, correspondingly, $\bar{B}$ entering in $\bar{M}$ is given by $\partial_\lambda \bar{A}$.

In practice, we again need to compute the rectangular matrix (\ref{deftildeM}).
For any $k^{\rm th}$-order excitation, $P\bar{M}P$ is a simple $k\times k$ square submatrix of $\tilde{M}$, which is built in the same way as $PA^{-1}\bar{A}P$ is built from $A^{-1}\tilde{A}$.
Then, one has to perform the trace of the inverse of the $k\times k$ matrix $PA^{-1}\bar{A}P$ times the matrix $P\bar{M}P$. The cost of this calculation is of order $k^3$  due to the computation of the inverse.
For example, if $1 \to 11$ and $3 \to 15$ are the list of excitations,
Eq.~(\ref{multiplex}) becomes 
\begin{equation}
\frac{\hat{O}\bar{\wdet}}{\bar{\wdet}} = \frac{\hat{O}\wdet}{\wdet} +
 \tr \left( \left[ \begin{matrix} (\aa)_{1,11} & (\aa)_{1,15}   \\ 
                      (\aa)_{3,11} & (\aa)_{3,15}                         
\end{matrix} \right]^{-1} \left[
\begin{matrix} \tilde{M}_{1,11} & \tilde{M}_{1,15}   \\ 
                      \tilde{M}_{3,11} & \tilde{M}_{3,15}                         
\end{matrix}  \right]
 \right)\,.
\end{equation}

For a mono-excitation $i \to j$, $\P$ projects on a one-dimensional space.  $(\P \A^{-1} \bar{\A}\P)^{-1}$ is therefore a scalar, which is equal to ${\wdet}/{\bar{\wdet}}$
thanks to Eq.~(\ref{detk}),  $\bar{M}$ is the matrix element $\tilde{M}_{ij}$, and $\bar{D}=\partial_\mu D$ where $\mu$ is a mono-excitation parameter introduced in equation  (\ref {devmono}).
Using Eq.~(\ref{multiplex}), we recover that for a mono-excitation
\begin{equation}
\tilde{M}_{ij} = \left(\frac{{\hat{O}}\bar{\wdet}}{\bar{\wdet}}-\frac{{\hat{O}}\wdet}{\wdet}\right)\frac{\bar{\wdet}}{\wdet} = \frac{d}{d\mu} \frac{{\hat{O}}\wdet}{\wdet}\,.
\end{equation}

We note that the first $N$ columns of $\tilde{M}$ are identically zero and, as in the case of the matrix $A^{-1}\tilde{A}$, do not have to be stored.
The matrix elements of $\tilde{M}$ 
should only be computed for the lines and columns corresponding to the $N_s$ active single excitations, which are in general fewer than the product $N_{\rm act}N_{\rm virt}$ of the
occupied active and virtual orbitals. 
The cost is $O(N^2 N_{\rm act})+ O(N_s N)$ if this product is evaluated from the left to the right, while it is $O(N^2 N_{\rm virt})+ O(N_s N)$ if one starts from the right.
 If the matrices $B$ and the corresponding $\tilde{B}$ are sparse, the cost is smaller. In particular, for the drift, these matrices have only one non-zero row and the additional cost of evaluating the derivative with respect to the coordinates of one electron is 
$O(N_s)$.

In Appendix~\ref{App_SM}, we also provide a more lengthy derivation of expression (\ref{multiplex}) using the Sherman-Morrison-Woodbury 
formula instead of performing the derivative of (\ref{detk}).

\section{Jastrow factor, pseudopotentials, and other expressions for $B$}
\label{jaspseB}

When a Jastrow factor is included, 
\begin{equation}
\psi({\bf R}) = J ({\bf R}) D ({\bf R}) = J ({\bf R}) \det (A({\bf R}))\,,
\end{equation}
and the expression of the matrix $B$ in a local quantity must be modified to account for the presence of the Jastrow factor.

We begin with the local kinetic energy,
\begin{equation}
\frac{\hat{T} \psi} {\psi}= -\frac{1}{2}\sum_i \frac{\Delta_i \psi}{\psi} = 
\frac{1}{\det(A)}\sum_i\left[-\frac{1}{2}\left(\Delta_i  + 2\frac{\nabla_i J}{J}\cdot {\nabla_i } + \frac{\Delta_i J}{J} \right)\right]\det(A)\,.
\label{Jkin}
\end{equation}
Following the derivations in Section~\ref{O_local}, we identify a generalization of the operator $O({\bf r}_i)$ as the operator within the square brackets and obtain 
\begin{equation}
B^{\rm kin}_{ij}= -\frac{1}{2} \left[
 \Delta \phi_j({\bf r}_i) + 2\frac{\nabla_i J}{J} \cdot \nabla \phi_j({\bf r}_i) + \frac{\Delta_i J}{J}  \phi_j({\bf r}_i) \right]\,.
\label{Bdef}
\end{equation}
The local kinetic energy can then be written as
\begin{equation}
\frac{\hat{T}    \psi}{\psi} =  \tr  (A^{-1} B^{\rm kin})\,.
\end{equation}

It is possible to cast the contribution of the potential to the local energy in a similar form,
starting from the more complicated non-local component
\begin{eqnarray}
\frac{\hat{V}_{\rm NL} \psi}{\psi}=\sum_i\sum_a\sum_l v_l^a(r_{ia})
\int_{|r^\prime_{ia}|=r_{ia}} \hspace*{-0.5em}d\Omega^\prime P_l(\cos\theta^\prime)\frac{\psi({\bf r}_1,\ldots,{\bf r}_i^\prime,\ldots,{\bf r}_N)}{\psi({\bf r}_1,\ldots,{\bf r}_i,\ldots,{\bf r}_N)}\,,
\end{eqnarray}
where the summations over $i$, $a$, and $l$ run over the electrons, the nuclei, and the angular components of the non-local pseudopotential, respectively. 
For each electron coordinate, the integral is over a sphere centered on a nucleus with radius given by the electron-nucleus distance
$r_{ia}=|{\bf r}_{ia}|=|{\bf r}_i-{\bf R}_a|$ and the angle $\theta^\prime$ is between the vectors ${\bf r}^\prime_{ia}$  and ${\bf r}_{ia}$.
In QMC, the integral is computed as a sum over quadrature points characterized by weights $w_q$ and 
unit directions $\hat{\bf u}_q$,
\begin{eqnarray}
\frac{{\hat{V}}_{\rm NL} \psi}{\psi}=\frac{1}{\det(A)}\sum_i\left[\sum_a\sum_l v_l^a(r_{ia})
\sum_{q} w_{q}P_l(\cos\theta^a_{q})\frac{J(\ldots,{\bf q}^a_i,\ldots)}{J(\ldots,{\bf r}_i,\ldots)}\right]\det(A(\ldots,{\bf q}^a_i,\ldots))
\label{vNL}
\end{eqnarray}
where ${\bf q}^a_i={\bf R}_a+r_{ia} \hat{\bf u}_q$ and $\theta^a_{q}$ is the angle between $\hat{\bf u}_q$ and ${\bf r}_{ia}$.
In general, a different number of angular components and quadrature points can be used for the different atom types.
We then identify the matrix $B^{\rm NL}$ as
\begin{eqnarray}
B^{\rm NL}_{ij}&=&\sum_a \sum_l v_l^a(r_{ia}) 
\sum_{q} w_{q}P_l(\cos\theta^a_{q})\frac{J(\ldots,{\bf q}^a_i,\ldots)}{J(\ldots,{\bf r}_i,\ldots)}\phi_j({\bf q}^a_i)\,,
\label{BNL}
\end{eqnarray}
so that
\begin{eqnarray}
\frac{{\hat{V}}_{\rm NL} \psi}{\psi}=\tr(A^{-1}B^{\rm NL})\,.
\end{eqnarray}
In analogy to the treatment of the Laplacian of the Jastrow in the local kinetic energy (Eq.~\ref{Jkin}), we can rewrite the contribution of the local potential as
\begin{eqnarray}
\hat{V}_{\rm loc} = \frac{1}{\det(A)} \sum_i\left[\sum_a v_{\rm loc}^a(r_{ia})+\sum_{j<i}\frac{1}{|{\bf r}_i-{\bf r}_j|}\right]\det(A)
\label{vL}
\end{eqnarray}
and define 
\begin{eqnarray}
B_{ij}^{\rm loc}=\left[ \sum_a v_{\rm loc}^a(r_{ia})+\sum_{j<i}\frac{1}{|{\bf r}_i-{\bf r}_j|} \right]\phi_{j}({\bf r}_i)\,.
\label{Bloc}
\end{eqnarray}

The complete matrix $B$ in the trace expression of the local energy is the sum of the kinetic (Eq.~\ref{Jkin}) and potential (Eqs.~\ref{BNL}~and~\ref{Bloc}) contributions.
From the rectangular extensions $\tilde{A}$ and $\tilde{B}$ of the $A$ and $B$ matrices to the unoccupied orbitals, one can compute the local energy of any multiple excitation at low cost and, therefore, of any wave function given by a CI expansion times a Jastrow factor,
\begin{equation}
\psi_{\rm CI} = J \left[\det (A) + \sum_I c_I \det (\bar{A}_I)\right]\,,
\end{equation}
where $A$ is the reference Slater matrix computed from the occupied orbitals and $\bar{A}_I$ is a $k_I^{\rm th}$-order excited Slater matrix. 
Once we have computed $B$ (Eq.~\ref{Bdef}), $A^{-1}$, $A^{-1}\tilde{A}$,  and $\tilde{M}$, the local energy reads 
\begin{eqnarray}
\frac{\hat{H}\psi_{\rm CI}}{\psi_{\rm CI}} = \tr (A^{-1}B) + \frac {\sum_I c_I \tr (\alpha_I^{-1} M_I) \det (\alpha_I)}{\sum_I c_I \det(\alpha_I)} 
\end{eqnarray}
where the matrices $\alpha_I$ and $M_I$ are $k_I \times k_I$ submatrices of $A^{-1}\tilde{A}$ and $\tilde{M}$, respectively.

In Appendix~\ref{App_force}, we give the expressions of the derivatives of the matrices $A$ and $B$ with respect to the atomic coordinates (in particular, the formula for the derivatives of $B^{\rm NL}$) needed in the computation of the interatomic forces. We also discuss how to efficiently evaluate the additional terms in the force estimator introduced by the use of the space-warp transformation on the electron~\cite{Umrigar89,Filippi00}, which also require the derivatives of $A$ and $B$ with respect to the electronic coordinates. From the extension of these matrices to the unoccupied orbitals, one can easily compute the forces for a general multi-determinant Jastrow-Slater wave function.

Before presenting numerical examples, we now discuss the computational cost of a typical QMC run using these formulas.
We assume that we compute the energy in either variational or diffusion Monte Carlo
after performing a full sweep over the electrons in an all-electron move or $N$ one-elecron moves, and give the total cost.
The evaluation of $\tilde{A}$, $A^{-1}$, and all needed $B$ matrices (3$N$ for the drift and one for the local energy) scales as $O(N^3 + N^2N_{\rm virt})$.
The total cost~\cite{act_virt} of the corresponding $\tilde{M}$  matrices is  
$O(N_{\rm virt}N^2 +  N_s N)$,
and the use of formula (\ref{multiplex}) for $N_e$ determinants O($N N_e$).  
The overall cost to build the sampling process is then $O(N^3) + O(N^2 N_{\rm virt})+ O (N N_s) + O(N N_e)$ where $N_s\leq N_{\rm act}N_{\rm virt}$. 
Note that the dependence on $N_s$ is $N$ times  smaller than the one presented in Ref.~\onlinecite{Clark11}, which is of course a significant gain when   $N_s$ is large.
Assuming now that $N_{\rm virt} \leq O(N)$ this scaling can be simplified to  $O(N^3) + O (N N_e)$ since $N_s \leq N_e$.

\section{Numerical examples}
\label{numexe}

We demonstrate the formulas above on the C$_n$H$_{n+2}$ molecular series with $n$=4-60. We employ the CHAMP code~\cite{Champ} with 
scalar-relativistic energy-consistent Hartree-Fock pseudopotentials and the corresponding cc-pVDZ basis set~\cite{Burkatzki07,Hpseudo}.
The Jastrow factor is limited to a simple two-body electron-electron term and the single determinant is built from Hartree-Fock 
orbitals.

\begin{figure}[h]
\includegraphics[width=\columnwidth]{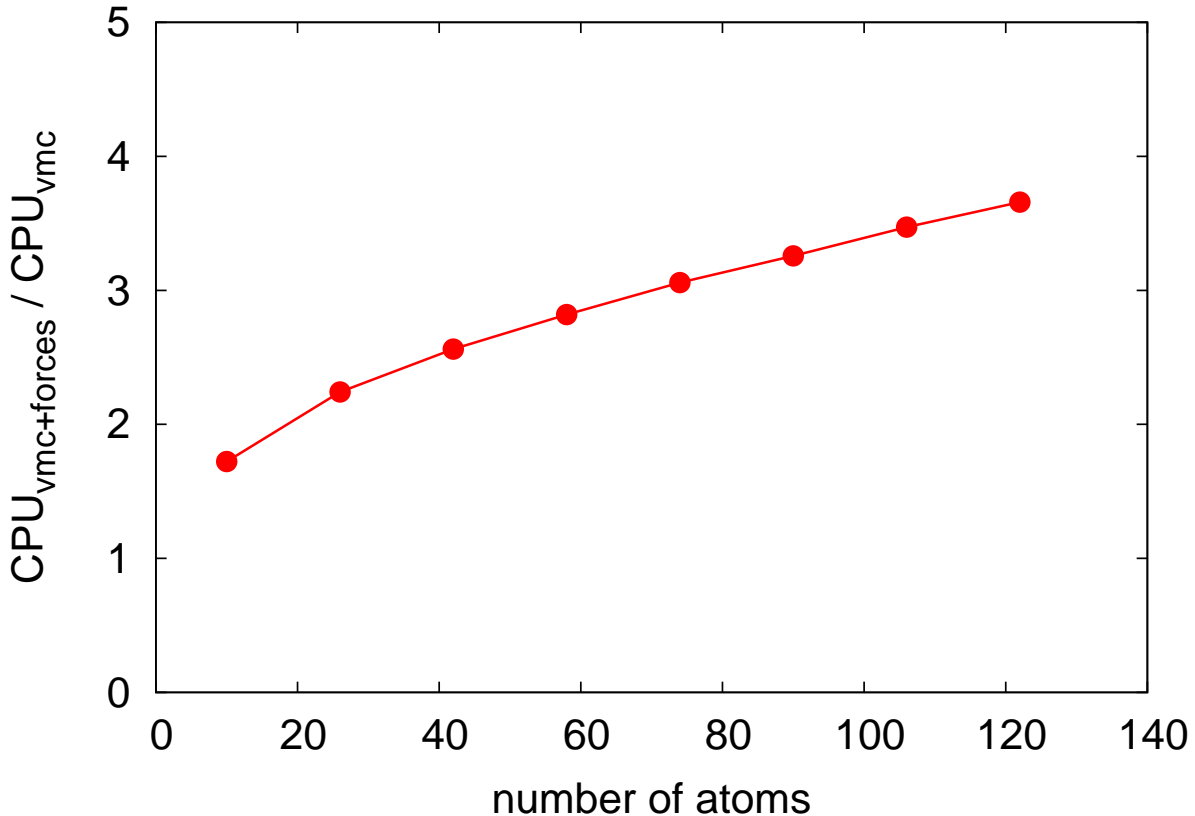}
\caption{
Ratio of the CPU time for a VMC calculation of the forces to the CPU time for the same 
simulation of the energy alone. The number of atoms refers to the sequence of molecules C$_n$H$_{n+2}$ with $n$ between 4 
and 60. The forces are calculated after moving all the electrons once.  
}
\label{forces_a}
\end{figure}

The low computational cost of the derivative expression (Eq.~\ref{dm_dettrace0}) in the VMC calculation of the 
interatomic forces for a one-determinant Jastrow-Slater wave function is demonstrated in Fig.~\ref{forces_a}:
For the largest system considered here which includes 122 atoms, computing all interatomic gradients costs less 
than 4 times a VMC simulation where one only evaluates the total energy.
A similar factor has been reported in Ref.~\onlinecite{Sorella10} where the forces were however evaluated with the aid of 
algorithmic differentiation (AD). Here, we demonstrate that a simple algebraic manipulation of the quantities needed to 
compute the forces leads to transparent, simple formulas to implement and an equivalent computational gain to the use of AD.
We note that the ratio of the CPU time of evaluating the local energy and the interatomic forces to the time of computing 
the energy alone should asymptotically be constant: The very weak linear dependence on the number of atoms (electrons) 
observed here is due to the $N^2$ term in the computational cost being more important in the energy than in the force calculation, 
at least at these system sizes.

\begin{figure}[h]
\includegraphics[width=\columnwidth]{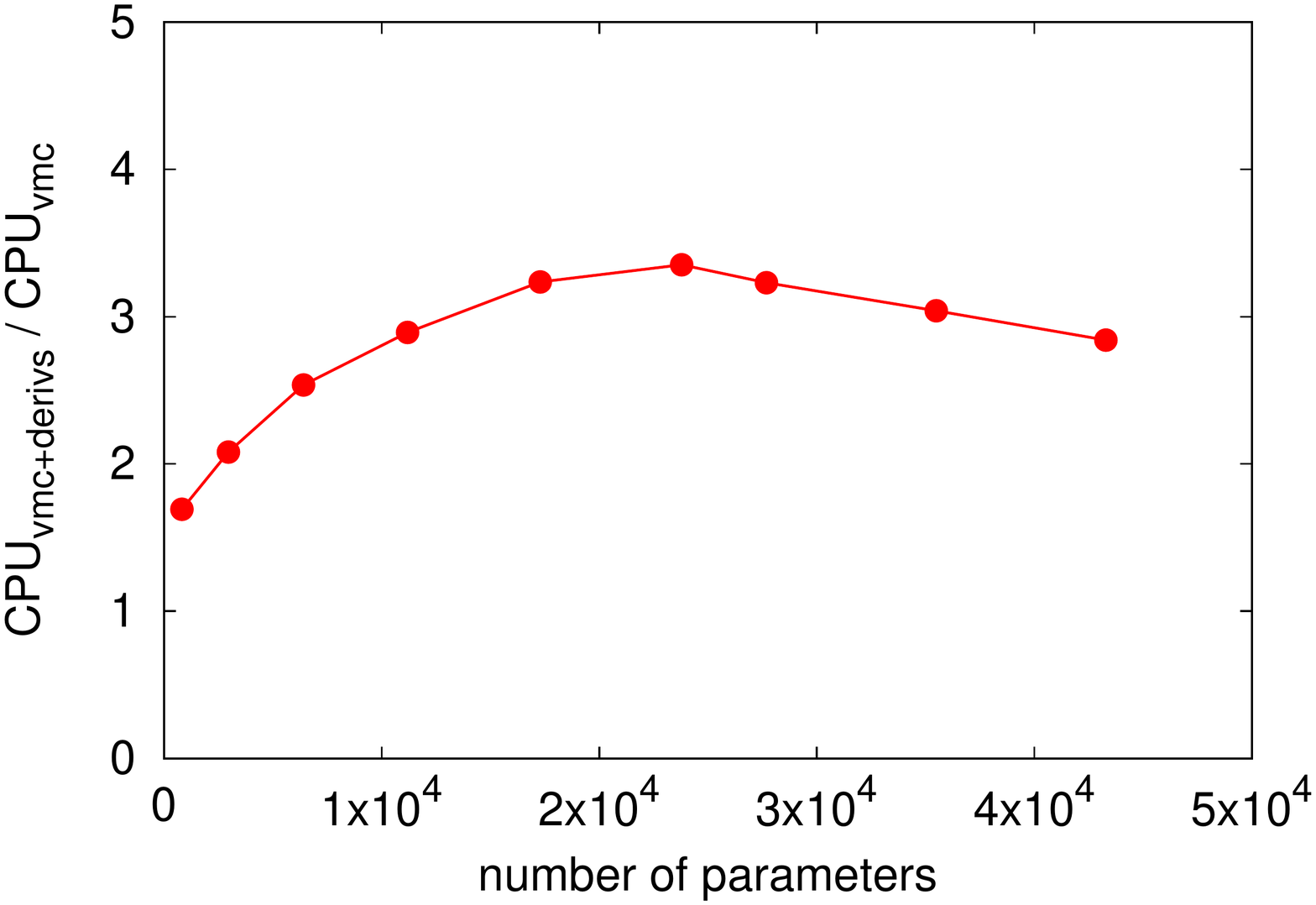}
\caption{
Ratio of the CPU time of a VMC simulation where all derivatives required for orbital optimization are computed,
and of the same simulation with the energy alone. The number of variational parameters refers to the sequence of 
molecules C$_n$H$_{n+2}$ with $n$ between 4 and 44. The derivatives are calculated after moving all the electrons once.  
}
\label{orb_a}
\end{figure}

We illustrate the gain achieved in the application of the same derivative expression to the orbital optimization of a one-determinant 
Jastrow-Slater wave function (Eq.~\ref{devmono}) in Fig.~\ref{orb_a}. For each system, we consider all possible orbital variations 
$\phi_i \to \phi_i+\mu\phi_j$ and compute the local energy together with the quantities $\partial_\mu\psi$ and 
$\hat{H}\partial_\mu \psi$ needed in the linear optimization method~\cite{Umrigar07}.  The ratio of the cost of such a VMC simulation
to the cost of only evaluating the local energy should not grow with system size: $\hat{H}\partial_\mu \psi$ can be straightforwardly 
obtained from $\partial_\mu (\hat{H}\psi/\psi)$ and the cost of calculating $\partial_\mu (\hat{H}\psi/\psi)$ for all possible orbital 
variations is proportional to $N^3$ as discussed in Section~\ref{2deriv_1det}. We find that this ratio remains well below 4 for system 
sizes leading to as many as $4.5\times 10^4$ orbital variations.

Finally, we demonstrate the speedup in a VMC simulation performed using expression (Eq.~\ref{multiplex}) to compute a local operator acting on a 
multi-determinant wave function. We focus on the local energy, which is evaluated after all the electrons have been moved once, and employ 
the same formula also in the computation of the gradient with respect to the coordinates of the electron being moved during the sweep over all 
the electrons.  For C$_4$H$_6$, C$_8$H$_{10}$, and C$_{16}$H$_{18}$, we generate a set of $N_e$ doubly excited determinants in either the up- or the down-spin 
component, treat all up- and down-spin determinants as distinct, and excite also from the core. Additional computational 
saving can therefore be achieved by exploiting that different excitations may share the same spin component~\cite{Scemama15} or by limiting the number 
$N_{\rm act}$ of active orbitals. For C$_{16}$H$_{18}$, we also investigate the use of triple excitations
in either the up- or the down-spin determinants.

\begin{figure}
\includegraphics[width=\columnwidth]{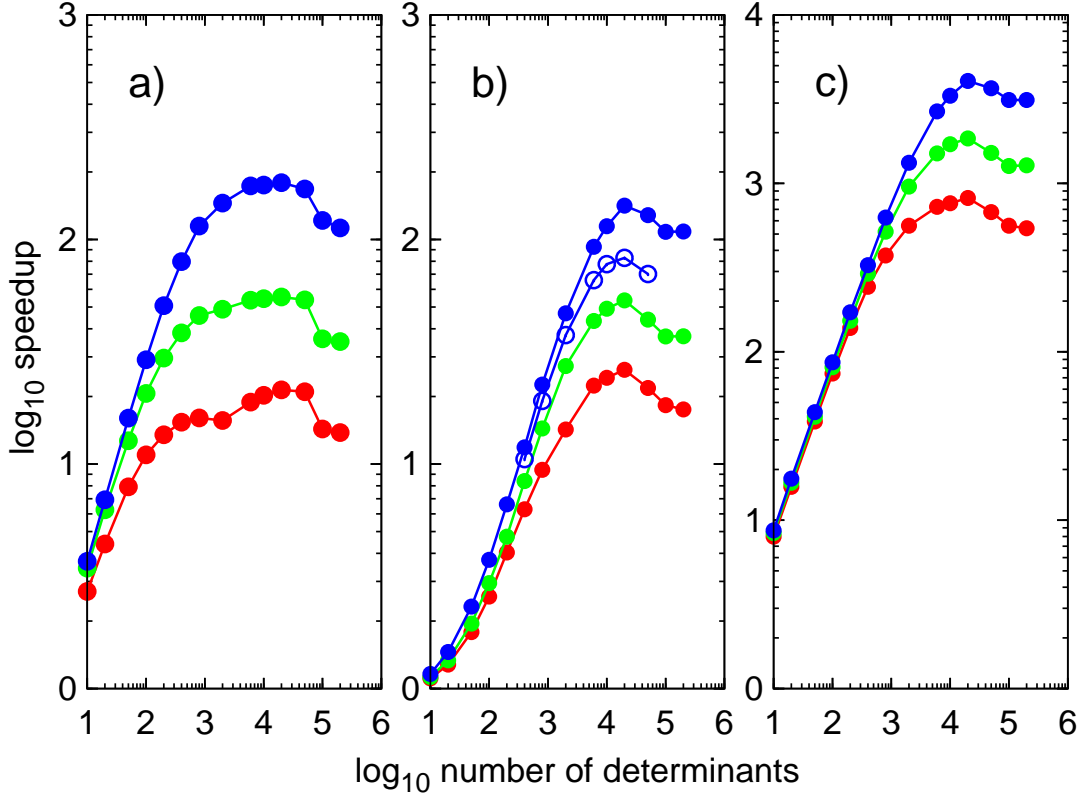}
\caption{
Speedup of the improved algorithm for a multi-determinantal wave function measured as a)
gain over a standard algorithm in the calculation of the wave function, the drift, and the relevant matrix updates,
b) gain over a standard algorithm in the complete VMC computation of the energy, and c) gain over $N_{\rm det}$ times 
the cost of a VMC run with one determinant.  The local energy is computed after a sweep over all electrons.
Data are shown for C$_4$H$_6$ (red), C$_8$H$_{10}$ (green),  and C$_{16}$H$_{18}$ (blue) as a function of the number of determinants. 
Filled (empty) circles refer to double (triple) excited determinants. All determinants in the expansion are treated as distinct and core
excitations are included. 
}
\label{mdet_a}
\end{figure}

In Fig.~\ref{mdet_a}, we present three different measures of speedup. In the left panel, we compare with the results presented in Ref.~\onlinecite{Clark11} (see
their Fig.~3 and green curve) and only estimate the cost of computing the wave function, the drift, and the relevant matrix updates (not the orbitals)
with respect to the standard method of computing and updating the inverse matrices of all determinants. In agreement with Ref.~\onlinecite{Clark11}, we find 
that the ratio between the two computational costs increases quickly with the number $N_{\rm det}$ of determinants before settling to a value, which we however
find to be greatly dependent on the used machine and compiler.
A more direct comparison would require further knowledge of the separate performance of the algorithms employed in the standard and improved calculation.
The speedup for more than approximately 100 determinants ranges between about 10 and 100 for the systems studied here.

The central panel represents a more realistic assessment of the formulas presented, showing the ratio of the time of a
complete VMC computation of the energy with the standard and the new algorithm. The speedup measured in this way is rather comparable to
what reported in the left panel, indicating that other parts of the code either scale similarly or do not affect the overall ratio.
As expected, the gain is a bit smaller for triple than for double excitations due to the larger dimension of the matrices needed to evaluate 
the local energy of a higher excitation (Eq.~\ref{multiplex}). 
Finally, the right panel allows a comparison with the way the speedup is measured in Ref.~\onlinecite{Scemama15}, where the cost of the
improved calculation is compared to $N_{\rm det}$ times the cost of a run with only one determinant, namely,
${N_{\rm det}\times{\rm time}(1)}/{{\rm time}(N_{\rm det})}$.
The higher values obtained with this measure, however, are not the speedups one gains in reality in comparison with the standard method, which is in fact 
much faster than just $N_{\rm det}$ times the cost of a simulation with a mono-determinantal wave function.

\section{Second-derivative of a $k^{\rm th}$-excited determinant}
\label{Sec_2nd_mult}

We compute here the derivative of a local quantity applied to an excited determinant $\bar{D}$. This is the equivalent of formula (\ref{dm_dettrace0}) applied to $\bar{D}$, and is obtained by differentiating expression (\ref{multiplex}) with respect to $\mu$,
\begin{equation}
\frac{\partial}{\partial \mu} \frac{\hat{O} \bar{D}}{\bar{D}} =\frac{\partial}{\partial \mu} \frac{\hat{O} D}{D} +
\tr \left((\partial_\mu  \alpha^{-1})P\bar{M}P + \alpha^{-1}(P\partial_\mu \bar{M}P )\right)\,,
\label{der2excit} 
\end{equation}
where $\alpha =  (P\A^{-1}\bar{\A}P)$. With the use of chain rule, it is straightforward to show that 
\begin{eqnarray}
\partial_\mu \tilde{M} = [A^{-1} \partial_\mu \tilde{B} -X \partial_\mu \tilde{A}]-[A^{-1} \partial_\mu B -X \partial_\mu A](A^{-1}\tilde{A}) -(A^{-1}\partial_\mu A) \tilde{M} 
\label{tildemudef}
\end{eqnarray}
and
\begin{equation}
\partial_\mu  (A^{-1}\tilde{A})  = A^{-1}\partial_\mu\tilde{A}- (A^{-1}\partial_\mu A)(\A^{-1}\tilde{A})\,,
\end{equation}
so that $\partial_\mu  \alpha^{-1} = -\alpha^{-1}P\partial_\mu  (A^{-1}\bar{A})P\alpha^{-1}$.
To evaluate these rectangular matrices, we need to extend the computation of $\partial_\mu A$ and $\partial_\mu B$ (Eq.~\ref{dm_dettrace0}) to the virtual orbitals while
other relevant matrices like $A^{-1}\tilde{A}$ and $\tilde{M}$ are already available from the computation of the excited determinant and the corresponding local quantity.
The matrices $P\partial_\mu \bar{M}P$ and $P\partial_\mu  (A^{-1}\bar{A})P$ are then simple $k\times k$ submatrices constructed from the elements whose row and column indices correspond to the substituted occupied and the excited orbitals, respectively.
We note that $\partial_\mu \tilde{M}$ is the matrix of second derivatives of the local quantity with respect to $\mu$ and the mono-excitation parameter (Eq.~\ref{devmono}).

It should be apparent by now that, in practice, one needs to calculate the product of $A^{-1}$ with other matrices as in $A^{-1}B$, $A^{-1}\tilde{B}$, $A^{-1}\partial_\mu A$, $A^{-1}\partial_\mu \tilde{A}$ etc.\ and that these matrix products constitute the building blocks of the second derivatives and all other quantities derived so far. Additionally, in the computation of the second derivatives, it might be computationally advantageous to evaluate the products $X\tilde{A}$, $X\partial_\mu A$, and $X\partial_\mu \tilde{A}$ as detailed above and in Section~\ref{2deriv_1det}. The formulas more explicitly written in terms of these products and therefore closer to the actual implementation are given for clarity in Appendix~\ref{App_not} .

Since the computational cost of building $\partial_\mu \tilde{M}$ is of order $N^2\times N_{\rm virt}$, the total cost of evaluating the last term in expression~(\ref{der2excit}) for a multi-determinant wave function typically becomes $O(N^2N_{\rm virt})+O(k^3 N_e)$ where $N_e$ is the number of excited determinants.  The same scaling will also characterize higher-order derivatives.
If $\mu$ represents the coordinate of one atom, the final cost to construct the gradient with respect to the $3 N_{\rm atoms}$ nuclear coordinates is
\begin{eqnarray}
  O(N^3)+  O (N^2 N_{\rm virt} N_{\rm atoms} ) + O (N_e N_{\rm atoms}) \,,
\nonumber
\end{eqnarray}
If the number of active occupied orbitals $N_{\rm act} < N_{\rm virt}$, the computation of the three terms in Eq.~\ref{tildemudef} can be carried out in the same way as discussed before~\cite{act_virt} for the matrix $\tilde{M}$. For a single-determinant wave function or a small expansion with $N_{\rm act}$ and/or $N_{\rm virt}$ small,  we recover the cost described in Section~\ref{2deriv_1det}.

Finally, we stress that the formula above describes not only the derivative of a local quantity but also the second derivative $\partial_\mu\partial_\lambda \ln \bar{D}$. The matrices $B$ and $\tilde{B}$ are then equal to $\partial_\lambda A$ and $\partial_\lambda \tilde{A}$, respectively. An expression equivalent to Eq.~\ref{der2excit} but where $\lambda$ and $\mu$ are treated on an equal footing is given in Appendix~\ref{App_not}.

\section{Concluding remarks}

We have presented general and simple formulas to efficiently compute derivatives of wave functions, local quantities, and their derivatives needed in QMC simulations, when 
the wave function $\psi$ is written as a Jastrow factor times an expansion of $N_e+1$ Slater determinants. 
 
The simplicity of the formulas stems from the fact that a derivative of one-determinant wave function and a local quantity 
such as the local energy are treated on an equal footing and expressed as a trace of matrices which only require the computation of one-body functions 
(molecular orbitals) and their derivatives. The extension of these formulas to excited determinants is straightforward: One evaluates the matrix elements also
for the virtual orbitals (unoccupied in the reference) and computes products involving the resulting rectangular matrices, sub-matrices and their 
inverses in the spirit of what Clark {\it et al.}~\cite{Clark11} had developed for the calculation of the multi-determinant wave function.
Furthermore, our formulation allows an easy generalization to higher derivatives. 
Regarding the efficiency, it leads to significant gains for large $N_e$ when one computes many local properties and/or derivatives in a practical simulation. 

\appendix

\section{Interatomic forces}
\label{App_force}

\subsection{Derivatives with respect to nuclei positions}

To calculate the derivative of the local energy and wave function with respect to the nuclear  coordinates,
we need to evaluate the corresponding matrices $dA$ and $dB$ in addition to the logarithmic derivatives of the Jastrow factor.  
The matrix elements of $dA$ are simply 
the derivative of the single-particle orbitals, which we expand on an atomic basis $\{\chi\}$ as
\begin{eqnarray}
A_{ij}=\phi_j({\bf r}_i)&=&\sum_a\sum_{l_a}^{L_a} b_{jl_a}\chi_{l_a}({\bf r}_i-{\bf R}_a)\,,
\label{orb}
\end{eqnarray}
where $L_a$ is the number of basis functions on atom $a$.  Then, we obtain 
\begin{eqnarray}
\nabla_aA_{ij}=\nabla_a\phi_j({\bf r}_i)=-\sum_{l_a}^{L_a} b_{kl_a}\nabla \chi_{l_a}({\bf r}_i-{\bf R}_a)
\end{eqnarray}
Consequently, since the gradients of the basis functions are also needed to calculate $\nabla_i\phi_j({\bf r}_i)$, the
computation of $dA$ only requires quantitites which are normally evaluated in sampling the local energy.
Similarly, the computation of $dB^{\rm kin}$ (Eq.~\ref{Bdef}) requires derivatives of the basis functions, most of which have
already been evaluated for the local energy, with the exception of the off-diagonal components of the hessian 
and the gradient of the laplacian of $\chi$.
We do not report here the relatively simple expression of $dB^{\rm kin}$ but focus on the somewhat more complicated $dB^{\rm NL}$ (Eq.~\ref{BNL}).

In taking the derivative of $B^{\rm NL}$ with respect to the nuclear coordinates, we need to consider its explicit dependence on the nuclear coordinates ${\bf R}_a$ as for instance in $\phi_j({\bf r})$ (Eq.~\ref{orb}), as well as the implicit dependence through the quadrature points. Therefore, we have
\begin{eqnarray}
&&\nabla_a B^{\rm NL}_{ij}=-\sum_q  w_{q} \phi_j({\bf q}^a_i) \frac{J({\bf q}^a_i)}{J({\bf r}_i)} \times\nonumber\\
&\times&\sum_l \left\{\left.\frac{d v_l^a(r)}{d r}\right|_{r=r_{ia}} \frac{{\bf r}_{ia}}{r_{ia}} P_l(\cos\theta_q^a)
+v_l^a(r_{ia})\left.\frac{dP_l(\cos\theta)}{d\cos\theta}\right|_{\cos\theta_q^a}
\left(\cos\theta_q^a\frac{{\bf r}_{ia}}{r_{ia}}-\hat{\bf u}_q\right)\frac{1}{r_{ia}} \right\}+\nonumber\\
&+&\sum_l v_l^a(r_{ia}) \sum_q w_q
P_l(\cos\theta_q^a)\left\{\left.{\nabla_i^\prime [\phi_j({\bf r}^\prime_i)J({\bf r}_i^\prime)]}\right|_{{\bf r}_i^\prime={\bf q}^a_i}
-\left.\nabla_i^\prime [\phi_j({\bf r}_i^\prime)J({\bf r}_i^\prime)]\right|_{{\bf r}_i^\prime={\bf q}^a_i}\cdot \hat{\bf u}_q \frac{{\bf r}_{ia}}{r_{ia}}\right\}\frac{1}{J({\bf r}_i)}\nonumber\\
                  &+&\sum_b\sum_l v_l^b(r_{ib}) 
\sum_{q} w_{q}P_l(\cos\theta_q^b)\nabla_a\left[\phi_j({\bf r}^\prime_i) \frac{J({\bf r}^\prime)}{J({\bf r}_i)}\right]_{{\bf r}_i^\prime={\bf q}^b_i}\,,
\label{grad_a_BNL}
\end{eqnarray}
where we simplified the notation as ${J({\bf r}^\prime_i)}/{J({\bf r}_i)}={J(\ldots,{\bf r}_i^\prime,\ldots)}/{J(\ldots,{\bf r}_i,\ldots)}$. Therefore, differentiating the term in the local energy  due to the non-local potential results in a very compact formula (instead of the multiple expressions presented in Ref.~\onlinecite{Badinski07}). This only requires the gradients of the orbitals and Jastrow factor with respect to the electronic and nuclear positions (Eq.~\ref{orb}) computed at the quadrature points, and simple quantities such as some geometrical terms and the derivatives of the radial components of the non-local potentials.

When the determinantal component of $\psi$ is a sum of determinants, we just need to compute the rectangular extensions 
$\nabla_a \tilde{A}$ and $\nabla_a \tilde{B}$ of the matrices $\nabla_a {A}$ and $\nabla_a {B}$.
Calculations of $\nabla_a (\ln \psi)$, and $\nabla_a {\rm E}_{\rm L}$  are then straighforward using expressions (\ref{multiplex}) and (\ref{der2excit}). 

\subsection{Warped coordinates}

An improved estimator of forces (and also other observables) is obtained through the use of warped coordinates~\cite{Umrigar89,Filippi00} which, as detailed in Refs.~\onlinecite{assaraf03,Sorella10}, introduces additional terms in the force estimator: For any component of the force, one also needs to compute ${\bf v}\cdot{\bf \nabla} (\ln \psi)$ and ${\bf v}\cdot{\bf \nabla}{\rm E}_{\rm L}$,  where the gradient is taken with respect to the $3N$ electron coordinates. 
The vector field ${\bf v}$ depends on the electron and nuclear positions, and is different for the force components of the different atoms.
These two terms can be written as first derivatives of $\ln \psi$ and ${\rm E}_{\rm L}$ 
\begin{eqnarray}
{\bf v}\cdot {\bf \nabla} \ln \psi ({\bf R}) & =&  \left.\frac {d }{d \mu} \ln \psi ({\bf R} + \mu {\bf  v}) \right|_{\mu =0}
\label{vgradpsi} \\
{\bf v}\cdot{\bf \nabla} {\rm E}_{\rm L} ({\bf R})  &=&  \left. \frac {d }{d \mu} {\rm E}_{\rm L} ({\bf R} + \mu {\bf  v}) \right|_{\mu =0}\,.
\label{vgradel}
\end{eqnarray}
When $\psi$ is a single determinant times a Jastrow factor $\psi = J \det (A)$, the first term (\ref{vgradpsi}) is 
\begin{equation}
{\bf v}\cdot  \frac{\nabla (J  \det(A))} { J \det(A)}= \tr (A^{-1} {\bf v}\cdot\nabla A) +   {\bf v}\cdot\frac{\nabla J}{J} 
\end{equation}
where the coefficients of the matrix ${\bf v}\cdot\nabla A$ are 
\begin{equation}
( {\bf v}\cdot\nabla A)_{ij} \equiv {\bf v}\cdot\nabla A_{ij} 
=  {\bf v}_i \cdot \nabla \phi_j({\bf r}_i)\,, 
\end{equation}
where, in the last term, only the 3 components corresponding to the $i$-th electron survive.
The second expression (\ref{vgradel})  is  the derivative  of the local energy ${\rm E}_{\rm L} = \tr (A^{-1} B)$, so
${\bf v}\cdot{\bf \nabla} {\rm E}_{\rm L}$ is given by the expression (\ref{dm_dettrace0}) with 
\begin{eqnarray}
\partial_\mu B & = & {\bf v}\cdot\nabla B \text{ \ \ where  }   ( {\bf v}\cdot\nabla B)_{ij} \equiv  {\bf v}\cdot\nabla B_{ij} \,.
\end{eqnarray}
We recall that, when pseudopotentials are employed, $B \equiv  {B}^{\rm kin} + {B}^{\rm NL}$
with ${B}^{\rm kin}$ and ${B}^{\rm NL}$ given in Eqs.~(\ref{Bdef}) and (\ref{BNL}). The expression of $\nabla {B}^{\rm NL}$ includes a subset of
the terms required to evaluate Eq.~\ref{grad_a_BNL}.

When the determinantal component of $\psi$ is a sum of determinants, we simply need to compute the rectangular extensions  
${\bf v}\cdot \nabla \tilde{A}$ and ${\bf v} \cdot \nabla \tilde{B}$ of the matrices  ${\bf v}\cdot \nabla {A}$ and ${\bf v} \cdot \nabla {B}$.
Again, calculations of (\ref{vgradpsi}) and (\ref{vgradel}) are then straighforward using expressions (\ref{multiplex}) and (\ref{der2excit}).
Note that, if we want to keep the calculation of ${\bf v}\cdot \nabla \tilde{A}$ and ${\bf v}\cdot \nabla \tilde{B}$ of order $O(N^2)$, the vector field 
${\bf v}$ should be localized around the atom whose force we are evaluating, i.e. ${\bf v}_i({\bf r}_i) = 0$ when the distance between the electron ${\bf r}_i$ and the atom we are considering is larger than a given threshold. This is in fact how the space-warp transformation was introduced in Refs.~\onlinecite{Umrigar89,Filippi00}.

\section{First derivative using  the Sherman-Morrison-Woodbury formula}
\label{App_SM}

We can also  obtain (\ref{multiplex}) using the Sherman-Morrison-Woodbury formula instead of performing the 
derivative of (\ref{detk}).  The calculation will be a bit longer and less straighforward, but allow us to better understand the relationship with the 
approaches followed in Refs.~\onlinecite{Clark11} and \onlinecite{Scemama15}.

We need to update the trace expression (\ref{derivtrace}) when the Slater matrix $A$ and its derivative $B$ are replaced by an excited Slater matrix $\bar{A}$ and its derivative $\bar{B}$, 
\begin{equation}
 \tr (\bar{A}^{-1}\bar{B})=  \tr (A^{-1}\bar{B} + (\bar{A}^{-1}-A^{-1}) \bar{B})\,.
\label{trab}
\end{equation}
Writing $\bar{A}=A + (\bar{A}-A)\P$ and applying the Sherman-Morrison-Woodbury formula, we have
\begin{eqnarray}
\bar{A}^{-1}-A^{-1} & =&  -A^{-1} (\bar{A}-A) (1+\P A^{-1} (\bar{A}-A))^{-1}\P A^{-1} \nonumber \\
& = &  -A^{-1} (\bar{A}-A) (1-\P  +\P A^{-1} \bar{A}\P)^{-1} \P A^{-1} \nonumber  \\
 & =&  -A^{-1} (\bar{A}-A) (\P A^{-1} \bar{A}\P)^{-1}A^{-1}
\label{SMW} 
\end{eqnarray}
Substituting this result in Eq.~\ref{trab}, we obtain 
\begin{eqnarray}
\tr (\bar{A}^{-1}\bar{B}) &= & \tr (A^{-1}\bar{B} - 
A^{-1} (\bar{A}-A)  (\P A^{-1} \bar{A}\P)^{-1}A^{-1}\bar{B})\nonumber   \\
& = & \tr (A^{-1}\bar{B} + 
(\P A^{-1} \bar{A}\P)^{-1}A^{-1}\bar{B}  -\bar{B} A^{-1} \bar{A}  (\P A^{-1} \bar{A}\P)^{-1}A^{-1})\,. \nonumber
\end{eqnarray}
In the last term in the trace,   writing  $\bar{B}=B +(\bar{B}-B)P$, expanding the product, and using that  
$\P A^{-1} \bar{A}\P (\P A^{-1} \bar{A}\P)^{-1}=P$, we obtain
\begin{eqnarray}
\tr (\bar{A}^{-1}\bar{B}) &= & \tr ( A^{-1}\bar{B}-(\bar{B}-B)PA^{-1}+
(\P A^{-1} \bar{A}\P)^{-1} (A^{-1}\bar{B}-A^{-1}BA^{-1} \bar{A})) \nonumber  \\
& = &  \tr ( A^{-1}B) + \tr
(\P A^{-1} \bar{A}\P)^{-1} (A^{-1}\bar{B}-A^{-1}BA^{-1} \bar{A}))\,,
\label{update}
\end{eqnarray}
where we used that $(\bar{B}-B)P = (\bar{B}-B)$ in the last line. This ends the proof.

Note that, in the method proposed by Scemama {\sl et al.}~\cite{Scemama15}, the inverses of all excited determinants are  updated 
with the Sherman-Morrison formula, which amounts to the restriction of (\ref{SMW}) to one-column updates \footnote{In this work,
a tree from the set of excited determinants is built such that a child Slater determinant differs from his parent in only one column.}. The left-hand 
side of  (\ref{trab}) is then computed straighforwardly.  The computational scaling of these updates is $O(N^2)$ per excited determinant, leading  
to an overall scaling of $O(N^3) + O(N^2 N_e)$ for a large number $N_e$  of excited determinants distinct in both spin components ($N_e \gg N$).
The formula  (\ref{update})   avoids  these updates for excited determinants  and  the scaling is reduced to  $O(N^3)+ O(N_e)$.

\section{Symmetric and more compact formulas,  higher order derivatives}
\label{App_not}

We rewrite the formulas (\ref{multiplex}) and (\ref{der2excit}) in a slightly different notation which well emphasizes how the building blocks
in our formulation are products of $A^{-1}$ with other matrices, and which is more convenient when handling higher-order derivatives.  As we 
had done in Eq.~\ref{2deriv_sym}, we express the formulas symmetrically in $\lambda$ and $\mu$ when second derivatives with respect to $\lambda$ and 
$\mu$ are considered.
Note that a straightforward implementation of these formulas is not necessarily the most efficient. In particular,
one should use the matrix $X$ like in (\ref{dm_dettrace0}),  when there are a large number of parameters $\mu$ as compared to $\lambda$ in a small determinantal expansion.

If the capital letter $C$ stands for a matrix depending on some parameters $\lambda$ and $\mu$, the corresponding lower-case letter $c$ indexed 
by $\lambda$ and $\mu$ is defined as
\begin{equation}
c_{\lambda \mu} \equiv A^{-1} \partial_{\lambda \mu} C =  A^{-1} \frac{\partial^2 C}{\partial \lambda \partial \mu}\,.
\label{cap2low} 
\end{equation}
This notation extends naturally to  derivatives of any order. 
For  example, $\tilde{a} \equiv A^{-1} \tilde{A}$  ($0^{\rm th}$ order), $\tilde{a}_\lambda \equiv A^{-1}\partial_\lambda \tilde{A}$, 
$a_\lambda \equiv A^{-1} \partial_\lambda A$ etc.  The matrices $\tilde{a}$, $\tilde{a}_\lambda$, $\tilde{a}_\mu$, and $\tilde{a}_{\lambda \mu}$ 
require the application of the inverse Slater matrix $A^{-1}$ to matrices of molecular orbitals and their derivatives. These 4 matrices are the 
basic quantities from which everything can be simply expressed.
For example  $a$, ${a}_\lambda$,  ${a}_\mu$, ${a}_{\lambda \mu}$ represent respectively the first N columns of these matrices. 
For  any order $n \geq 0$, the following algebraic identity holds
\begin{equation} 
\partial_{\alpha} c_{\alpha_1 \dots \alpha_n}  =     c_{\alpha \alpha_1 \dots \alpha_n} -a_\alpha {c_{\alpha_1 \dots \alpha_n}}\,.
\label{derc}
\end{equation}
With this formula, the derivative of the rectangular extension of the slater matrix becomes
\begin{equation}
\partial_\lambda \tilde{a} =  \tilde{a}_\lambda -a_\lambda \tilde{a}\,,
\label{der1tilde}
\end{equation}
which is the matrix $\tilde{M}$ given in Eq.~\ref{deftildeM}.
The second-order derivative is 
\begin{eqnarray} 
\partial_{\lambda \mu} \tilde{a} & =&  \tilde{a}_{\lambda \mu} -a_\mu \tilde{a}_\lambda  \nonumber \\
  &-& a_\lambda (\tilde{a}_\mu -a_\mu \tilde{a}) \nonumber \\
&-&({a_{\lambda \mu}-a_\mu a_\lambda})\tilde{a} \nonumber \\
 & = &  \tilde{a}_{\lambda \mu}+(-{a_{\lambda \mu}+a_\mu a_\lambda+a_\lambda a_\mu})\tilde{a} - a_\lambda \tilde{a}_\mu- a_\mu \tilde{a}_\lambda.
\label{der2tilde}
\end{eqnarray} 
It is a symmetric expression in the parameters $\lambda$ and $\mu$ of the matrix $\partial_\mu \tilde{M}$ given in Eq.~\ref{tildemudef}.

For a given $k^{th}$-order excitation of the original Slater matrix $A$, we remind that $P\bar{a}P = PA^{-1} \bar{A}P$ is a $k\times k$ square submatrix  of $\tilde{a} = A^{-1} \tilde{A}$.
Every quantity introduced in this paper depends on the logarithmic derivatives of  
\begin{equation}
\det (\bar{A}) = \det(A) \det_P (PA^{-1}\bar{A}P) =  \det (A) \det_P (P\bar{a}P)\,.
\end{equation}
The first order derivative is 
\begin{eqnarray}
\partial_\lambda \ln (\det (\bar{A})) 
 &= &   \tr(a_\lambda) +\tr ((P\bar{a}P)^{-1} P \partial_\lambda \bar{a} P )\,.
\end{eqnarray}
The matrix $P \partial_\lambda \bar{a} P$ is a square submatrix of the matrix $\tilde{M} =\partial_\lambda \tilde{a}$. 
The same property holds for a derivative of any order: For any list $l=(\alpha_1 \dots \alpha_n)$ of derivation parameters, $P \partial_l \bar{a} P$ is a square submatrix of the matrix $\partial_l \tilde{a}$.
In analogy to (\ref{cap2low}), we introduce the notation  
\begin{equation}
p_{l} \equiv (P\bar{a}P)^{-1} P(\partial_{l} \bar{a}) P\,.
\label{defp}
\end{equation}
The same algebraic identity as (\ref{derc}) holds
\begin{equation}
\partial_{\alpha} p_l  =     p_{\alpha l} -p_\alpha {p_l}\,.
\end{equation}
With this  notation, the first derivative is
\begin{equation}
\partial_\lambda \ln (\det (\bar{A}))  =  \tr(a_\lambda) +\tr (p_\lambda)\,,
\end{equation}
and the second derivative is
\begin{eqnarray}
\partial_{\mu \lambda} \ln (\det (\bar{A})) & =&  \tr(a_\lambda-a_\mu a_\lambda )+\tr \left(p_{\lambda \mu} -p_\mu p_\lambda \right) .
\end{eqnarray}
We can also write the third order derivative
\begin{eqnarray}
\partial_{\alpha \mu \lambda} \ln (\det (\bar{A})) & = &\tr \left(a_{\alpha \lambda \mu}     + (- a_{\lambda \mu} + a_{\mu} a_\lambda +a_\lambda a_\mu)a_\alpha   - a_{\lambda \alpha}a_{\mu  } -a_{\mu \alpha } a_\lambda \right) \nonumber \\ 
& + &\tr \left(p_{\alpha \lambda \mu}     + (- p_{\lambda \mu} + p_{\mu} p_\lambda +p_\lambda p_\mu)p_\alpha   - p_{\lambda \alpha}p_{\mu  } -p_{\mu \alpha } p_\lambda
 \right)
\end{eqnarray}
In all these expressions,  the first trace is the logarithmic derivative of the original determinant (occupied orbitals) and the second trace 
is the  corrective term for a $k^{th}$-order excitation. Both terms  have exactly the same algebraic structure. 

The $k \times k$ matrices $p_l$ are easy to compute: One uses expression (\ref{defp}) and notes that the matrix $P\partial_\lambda \bar{a}P$ is a 
simple square submatrix of $\partial_l \tilde{a}$, which is itself given by the recursion formula (\ref{derc}) or expressions (\ref{der1tilde}) and
(\ref{der2tilde}).  All these quantities depend only on $\tilde{a}=A^{-1}\tilde{A}$ and the related transformed derivatives 
$\tilde{a}_l=A^{-1}\partial_l \tilde{A}$.  Writing expressions for higher-order derivatives is straighforward.

\acknowledgments
We thank Bryan Clark and Miguel Morales for useful discussions.
C.F.\ acknowledges support from an ECHO grant (712.011.005) of the Netherlands Organisation for Scientific Research (NWO).
This work was supported in part by the Italian MIUR through PRIN 2011.

\bibliography{paper.bib}

\begin{thebibliography}{15}
\bibitem [{act()}]{act_virt}%
  \BibitemOpen
  \href@noop {} {}\bibinfo {note} {If the products are carried out from the
  left to the right, the factor $N_{\rm virt}$ appearing in the scaling must be
  substituted with $N_{\rm act}$.}\BibitemShut {Stop}%
\bibitem [{Hps()}]{Hpseudo}%
  \BibitemOpen
  \href@noop {} {}\bibinfo {note} {For the hydrogen atom, we use a more
  accurate BFD pseudopotential and basis set. Dolg, M.; Filippi, C., private
  communication}\BibitemShut {NoStop}%
\end{thebibliography}

\end{document}